\documentclass[a4paper,11pt]{article}
\usepackage{epsfig}

\newcommand{\bea}{\begin{eqnarray}}
\newcommand{\eea}{\end{eqnarray}}
\newcommand{\be}{\begin{equation}}
\newcommand{\ee}{\end{equation}}

\begin{document}

\setcounter{page}{0}
\thispagestyle{empty}

\begin{flushright}
May 2001 \\
KIAS P01030 \\
\end{flushright}

\vspace{2cm}

\begin{center}
{\large \bf 
Probing large extra dimensions with spin configuration
of top quark pair production at the JLC
\footnote{To be published in the proceedings of the theory meeting 
on physics at Linear Colliders, 15 -- 17 March 2001, KEK, Japan}
}

\end{center}
\baselineskip=32pt

\begin{center}
Kang Young Lee, JeongHyeon Song\\
\baselineskip=22pt
{\footnotesize \it 
School of Physics,
Korea Institute of Advanced Study,
Seoul 130-012, Korea} \\
\baselineskip=28pt
Seong Chan Park, H. S. Song, Chaehyun Yu\\
\baselineskip=22pt
{\footnotesize \it 
Department of Physics,
Seoul National University, Seoul 151-742, Korea
}
\end{center}

\vspace{1cm}

\begin{abstract}

We explore signatures of large extra dimensions 
with a polarized electron/positrn beam
at the Joint Linear Collider (JLC).
We point out that spin informations can be useful 
to study indirect signals for large extra dimensions 
due to spin-2 nature of Kaluza-Klein gravitons.
The spin configurations of the top quark pair production
at $e^+ e^-$ and $\gamma \gamma$ collisions
provide an unique testing ground to search for 
effects of the large extra dimensions.
Especially we show the forward-backward asymmetry
is a good probe at $e^+ e^- \to t \bar{t}$ process
and we can define a new asymmetry observable 
effective for $\gamma \gamma \to t \bar{t}$ process.
\end{abstract}
\vspace{2cm}
\vfill
\pagebreak

\normalsize\baselineskip=15pt
\setcounter{footnote}{0}
\renewcommand{\thefootnote}{\arabic{footnote}}

\newcommand{\tabtopsp}[1]{\vbox{\vbox to#1{}\vbox to1{}}}
\newcommand{\gsim}{\buildrel>\over{_\sim}}
\newcommand{\lsim}{\buildrel<\over{_\sim}}
\renewcommand{\thefootnote}{\fnsymbol{footnote}}
\newcommand{\tchi}{\tilde{\chi}}
\newcommand{\psla}{p\kern-.45em/}
\newcommand{\esla}{E\kern-.45em/}

\section{Introduction}

Recently Arkani-Hamed, Dimopoulos, and Dvali (ADD) \cite{ADD1}
have suggested a model where the size of extra dimensions
could be large enough to be detectable
if we confine the matter fields
to our 4-dimensional world.
According to this idea,
the weakness of the gravity in our world
originates in the suppression arising from the
large volume of extra dimensions 
since graviton still propagates freely 
in the whole $(4+N)$-dimensional spacetime.
Considering the macroscopic Gauss' law for the Newtonian gravity,
the 4-dimensional Planck scale is simply related to 
the fundamental scale of nature, called $M_S$,
of which the value is comparable to the electroweak scale
via the volume factor as
\begin{equation}
M_{Pl}^2 \sim M_S^{N+2} R^N.
\end{equation}
where $R$ is the typical size of an extra dimension.
Thus the hierarchy of the Planck scale $M_{Pl}$
and the electroweak scale $M_{_W}$ is reduced to
the revelation of the effects of the large extra dimensions
in the ADD scinario.
The $N=1$ case is excluded by this simple relation
because the corresponding size of the extra dimension
is of order $10^{13}$ cm
when $M_S$ is at a few TeV order.
The $N=2$ case implies mm scale extra dimensions
and is very interesting in the view of macroscopic gravity
because it is not excluded by the current measurement of
gravitational force yet \cite{macro}.
However this case suffers from the strict constraint
from astrophysical arguements \cite{astro}.  
No other serious constraints exists up to now
for the cases of $N>2$.

The detectibility at high energy colliders is one of 
the most attractive aspects
of the large extra dimension idea.
When the momentum of gravitons involved in the process
does not exceed $M_S$,
the spacetime can be approximately described
by the linear expansion around the flat metric 
and we can derive
an effective action in the 4-dimensional spacetime
by compactifying the extra dimensions,
which leads to corresponding Feynman rules \cite{han,giudice}.
The Kaluza-Klein (KK) reduction
from the whole $(4+N)$-dimension to our 4-dimension
yields the 4-dimensional effective theory
involving towers of massive KK states
with cutoff at the string scale $M_S$.
Each KK state
interacts with the ordinary matter fields
with the couplings suppressed by the Planck scale.
However, the production of a single graviton is enhanced 
by the kinematic factor and has been studied 
as a source of the missing energy
in various processes \cite{giudice,Peskin}.
The indirect effects of massive graviton exchange
are enhanced by the sum of the tower of the KK states
and provide various signals in the collider phenomenologies 
[7-15].
In particular the spin-two nature of the gravitons
will result in the characteristic effects on the polarization observables
[8-10].

In this work, we explore effects of large extra dimension 
on the spin configuration in $t \bar{t}$ production.
In section II,
We suggest that the spin configuration
of top quark pair provide a promising signature 
to prove large extra dimensions in the top quark pair production
of $e^+ e^- $ collisions at the proposed Joint Linear Collider (JLC).
Further study at the $\gamma \gamma$ collider is performed
in section III. 
We summarize our results and conclude in section IV.

\section{Spin configuration of the top quark pair at $e^+ e^-$ colliders}

The produced top quark pair at the polarized $e^+ e^-$ collider 
is known to be in a unique spin configuration \cite{parke1}.
Since the lifetime of the top quark is too short to constitute hadrons,
information of the top polarization, which is not lost through hadronization,
can be read out through the angular distribution of the decay products 
\cite{spin-correlation}.
The neutral current nature of the graviton interactions
leaves the electroweak decay of top quark intact,
implying that the Standard Model (SM) prediction of the
angular correlations between the decay products and
the spin orientation of each top quark is still valid. 
Therefore the spin configuration of the top quark pair
can be a good probe of the effects of KK gravitons
at the top quark pair production process \cite{leetop}.

For the process
\begin{equation}
e^-(k_1) + e^+(k_2) \rightarrow t(p_1) + \overline{t}(p_2)
\;,
\end{equation}
the scattering amplitude of the $s$-channel Feynman diagram
mediated by the spin two gravitons summed over the KK tower
can be written by
\begin{eqnarray}
\label{top-amp}
{\cal M}_G &=& \frac{\lambda}{M_S^4} \Big[  (k_1-k_2)\cdot(p_1-p_2)
   \overline{v}(k_2) \gamma_\mu u(k_1) \overline{u}(p_1)\gamma^\mu v(p_2)
    \nonumber \\
     & &~~~~~~~~~ + \overline{v}(k_2)(p_1\hspace{-3.5mm}/
                     - p_2 \hspace{-3.5mm}/ \hspace{2mm})
      u(k_1) \overline{u}(p_1)(k_1\hspace{-4mm}/ - k_2\hspace{-4mm}/
      \hspace{2mm}) v(p_2) \hspace{2mm}\Big],
\end{eqnarray}
where the order one parameter $\lambda$ is assumed to include
the detailed structure of quantum gravity
such as the number of extra dimensions
and the compactification models.
Hereafter $\lambda=\pm1$ cases are to be considered,
which are sufficient for an estimation of the scale $M_S$.
It is to be noted that the amplitude in Eq. (\ref{top-amp})
as well as the SM amplitudes at the tree level
are CP invariant.

In order to analyze the spin configuration of the top quark pair,
let us briefly review a generic spin basis discussed in Ref. \cite{parke1}.
We define the spin states of the top quark and top anti-quark
in their own rest frame by decomposing their spins
along the reference axes $\hat{\eta}$ and $\hat{\bar{\eta}}$,
respectively.
The CP invariance, which is valid at the tree level 
even with the large extra dimension effects,
does not allow the T odd quantity, $i.e.$,
$\vec{\sigma}_t \cdot (\vec{k_1}|_{(t{\rm -rest})} \times
\vec{p_2}|_{(t{\rm -rest})})$
where the $\vec{\sigma}_t$ is the spin of the top quark,
and $\vec{k_1}|_{t{\rm -rest}}$ and $\vec{p_2}|_{t{\rm -rest}}$
are the momenta of the electron and the top anti-quark
in the rest frame of the top \cite{parke1,Kane}.
Thus the top and anti-top spins are to lie in the production plane.
The spin four-vectors of the top quark pair are chosen to be
back-to-back in the zero momentum frame.
The $\hat{\eta}$ is expressed in terms of an angle $\xi$ 
between $\hat{\eta}$ and the top anti-quark momentum 
in the rest frame of the top quark.
The familiar helicity basis is obtained by taking $\xi=\pi$.

In this general spin basis the differential cross sections
of the $e^+ e^- \to t\bar{t}$ process
with the large extra dimension effects are \cite{leetop}
\begin{eqnarray}
\displaystyle
\frac{d\sigma}{d\cos\theta}
  (e^-_Le^+_R\rightarrow t_\uparrow \bar{t}_\uparrow ~{\rm or}~
  t_\downarrow \bar{t}_\downarrow) &=&
   \frac{N_c\pi\alpha^2\beta}{2s}
  |\tilde{A}_{L}\cos\xi-\tilde{B}_{L}\sin\xi|^2 ,\nonumber \\
\frac{d\sigma}{d\cos\theta}
  (e^-_Le^+_R\rightarrow t_\uparrow \bar{t}_\downarrow ~{\rm or}~
  t_\downarrow \bar{t}_\uparrow) &=&
   \frac{N_c\pi\alpha^2\beta}{2s}
  |\tilde{A}_{L}\sin\xi+\tilde{B}_{L}\cos\xi\pm\tilde{D}_{L}|^2,\nonumber \\
\frac{d\sigma}{d\cos\theta}
  (e^-_Re^+_L\rightarrow t_\uparrow \bar{t}_\uparrow ~{\rm or}~
  t_\downarrow \bar{t}_\downarrow) &=&
   \frac{N_c\pi\alpha^2\beta}{2s}
  |\tilde{A}_{R}\cos\xi-\tilde{B}_{R}\sin\xi|^2 ,\nonumber\\
\label{general}
\frac{d\sigma}{d\cos\theta}
  (e^-_Re^+_L\rightarrow t_\uparrow \bar{t}_\downarrow ~{\rm or}~
  t_\downarrow \bar{t}_\uparrow) &=&
   \frac{N_c\pi\alpha^2\beta}{2s}
  |\tilde{A}_{R}\sin\xi+\tilde{B}_{R}\cos\xi\mp\tilde{D}_{R}|^2,
\end{eqnarray}
where $t_\uparrow\;(t_\downarrow)$
denotes the top spin along (against) the $\hat{\eta}$,
$N_c$ is the number of color, $\alpha$ is the fine structure
constant, $\beta=\sqrt{1-{4m_t^2}/{s}}$ and
\begin{eqnarray}
\tilde{A}_{L} &=& \frac{1}{2}(f_{LL}+f_{LR})
                \sin\theta\sqrt{1-\beta^2}-f_G\sin 2\theta
                \sqrt{1-\beta^2} ,\nonumber \\
\tilde{B}_{L} &=& \frac{1}{2}\bigg[f_{LL}(\cos\theta+\beta)
                +f_{LR}(\cos\theta-\beta)\bigg]
                -f_G\cos2\theta
                ,\nonumber \\
\tilde{D}_{L} &=& \frac{1}{2}\bigg[f_{LL}(1+\beta\cos\theta)
       +f_{LR}(1-\beta\cos\theta)\bigg] -f_G\cos\theta
        ,\nonumber \\
\tilde{A}_{R} &=& \frac{1}{2}(f_{RR}+f_{RL})\sin\theta\sqrt{1-\beta^2}
                -f_G \sin 2\theta\sqrt{1-\beta^2} ,\nonumber \\
\tilde{B}_{R} &=& \frac{1}{2}\bigg[f_{RR}(\cos\theta+\beta)
       +f_{RL}(\cos\theta-\beta)\bigg]
        -{f_G}\cos2\theta
                ,\nonumber \\
\tilde{D}_{R} &=& \frac{1}{2}\bigg[f_{RR}(1+\beta\cos\theta)
       +f_{RL}(1-\beta\cos\theta) \bigg]
        -{f_G}\cos\theta
.
\end{eqnarray}
The large extra dimension
effects are altogether included in the quantity $f_G$ defined by
\begin{equation}
\label{fG}
f_G = \frac{\beta s^2}{4\alpha} \frac{\lambda}{M_S^4}
\;.
\end{equation}
Here $f_{IJ}$'s $(I,J=L~ {\rm or}~ R)$ are
\begin{equation}
f_{IJ} = Q_\gamma(e)Q_\gamma(t)+Q_Z^I(e)Q_Z^J(t)\frac{1}{\sin^2\theta_W}
      \frac{s}{s-M_Z^2},
\end{equation}
and $\theta$ is the scattering angle of the top quark
with respect to the electron beam.
The electric charges and couplings to the $Z$ boson
of the electron and the top quark are given by
\begin{eqnarray}
Q_\gamma(e)=-1,
&Q_Z^L(e)=\frac{2\sin^2\theta_W-1}{2\cos\theta_W},&
Q_Z^R(e)=\frac{\sin^2\theta_W}{\cos\theta_W},
\nonumber\\
Q_\gamma(t)=\frac{2}{3},
&Q_Z^L(t)=\frac{3-4\sin^2\theta_W}{6\cos\theta_W},&
Q_Z^R(t)=-\frac{2\sin^2\theta_W}{3\cos\theta_W}.
\end{eqnarray}

\begin{figure}[t]
\epsfig{file=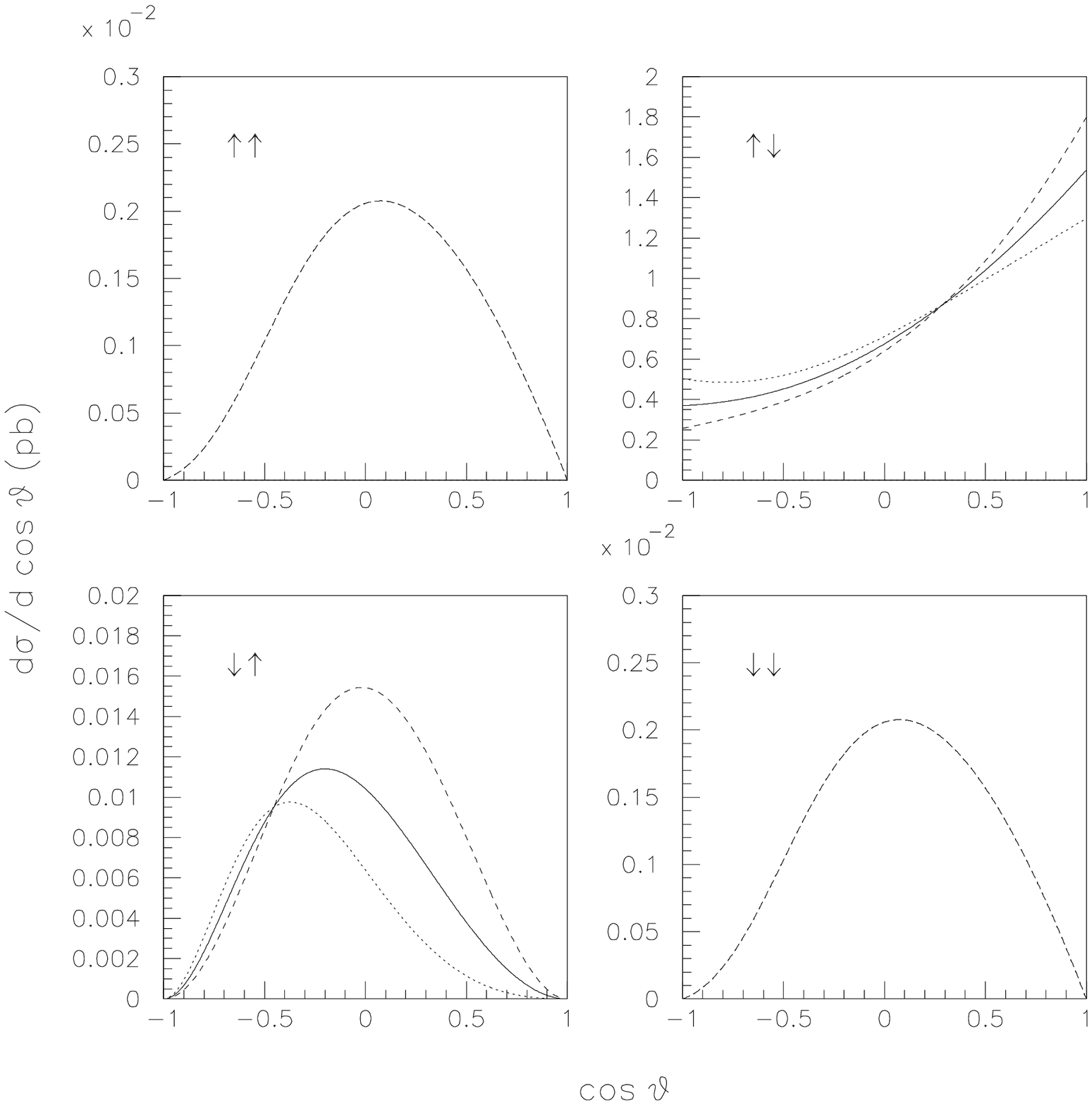,height=10cm,width=11cm}
\vspace{0.2cm}
\begin{center}
{\footnotesize Fig. 1 : The differential cross section with respect
to the scattering angle of the top quark
at  $\sqrt{s}=500$ GeV with the left-handed electron beam,
broken down to the spin configuration of the top quark pair.
The dotted (dashed) line includes the large extra dimension effects
when $\lambda=+1$ $(\lambda=-1)$ and $M_S=2$ TeV.
The solid line denotes the SM background.
}
\end{center}
\end{figure}

There exist the angles $\xi_{L}$ and $\xi_{R}$ such that the
differential cross sections for the
$t_\uparrow \bar{t}_\uparrow$ and $t_\downarrow \bar{t}_\downarrow$,
{\it i.e.,} like-spin states
vanish for the left- and right-handed electron beam, respectively.
It is called the ``{\it off-diagonal basis}",
of which the name originated in this feature \cite{parke1}.
In the SM, the angles  $\xi_{L,R}$ are taken to be
\begin{equation}
\label{xi}
\cos\xi_I = -\frac{B_I}{\sqrt{A_I^2+B_I^2}}\;,
\quad
\sin\xi_I = -\frac{A_I}{\sqrt{A_I^2+B_I^2}}\;,
\end{equation}
where $I=L,R$, $A_I=\tilde{A}_I|_{f_G=0}$, and $B_I=\tilde{B}_I|_{f_G=0}$.
There are two characteristic features of the SM
predictions in the off-diagonal basis.
First, the differential cross sections
for the like-spin states of the top quark pair vanish
or we have chosen the spin configuration in that way.
Second, the process for $t_\uparrow \bar{t}_\downarrow$
$(  t_\downarrow \bar{t}_\uparrow)$ is dominant when the
left-handed (right-handed) electron beam is used.
At high energy, the degree of this dominance is close to
100 $\%$ \cite{Why}.
This pure dominance of the up-down state for the left-handed
electron beam and the down-up state for the right-handed one
is fairly stable by the one-loop QCD corrections where
the soft gluon emissions are dominant so that the QCD corrections
are factored out.

The TeV scale quantum gravity modifies these two features.
First, the differential cross sections of the like-spin states
acquire contributions from quantum gravity.
Secondly the presence of $f_G$ in the differential cross sections of the
up-down and down-up states pollutes their pure dominance.
However the latter has two sides,
being weakened and strengthened dependent upon the sign of $\lambda$.

In Fig.~1 we plot the differential cross sections with respect to the
top quark scattering angle,
broken down to the spin configuration of the top quark pair
at $\sqrt{s}=500$ GeV with the left-handed electron beam.
We investigate that the like-spin states,
which are zero in the tree level SM, gain sizable contributions.

We observe that the angular distribution of the
cross sections provides valuable information
on the nature of the interactions between gravitons and fermions.
According to the sign of the $\lambda$,
the quantum gravity corrections act in a different way.
In the dominant processes (the up-down spin state with
the left-handed electron beam and the down-up spin state with the
right-handed one),
the virtual graviton exchanges cause destructive (constructive)
interference with the SM diagrams to the backward direction
when $\lambda=+1$ $(\lambda=-1)$.
To the forward direction,
on the contrary,
constructive (destructive) interference occurs
when $\lambda=+1$ $(\lambda=-1)$.
In the next dominant processes
(the down-up spin state with $e^-_L$
and the up-down one with $e^-_R$),
large quantum corrections with $\lambda=+1$
to the backward direction are reduced and dispersed.
Therefore we suggest that the forward-backward asymmetry $A_{FB}$
be very effective to probe
the TeV scale quantum gravity corrections.
Figure 2 illustrates the $A_{FB}$ for each sign of $\lambda$
with respect to $M_S$.
Assuming the 1 \% error, including systematic and statistical errors,
we expect the reach of bound on $M_S$ as 2.9 TeV.
Moreover, this observable clearly discriminates the $\lambda=-1$ case
from the $\lambda=+1$ case,
as their deviations go to the opposite directions from
the SM prediction.

\begin{figure}[t]
\epsfig{file=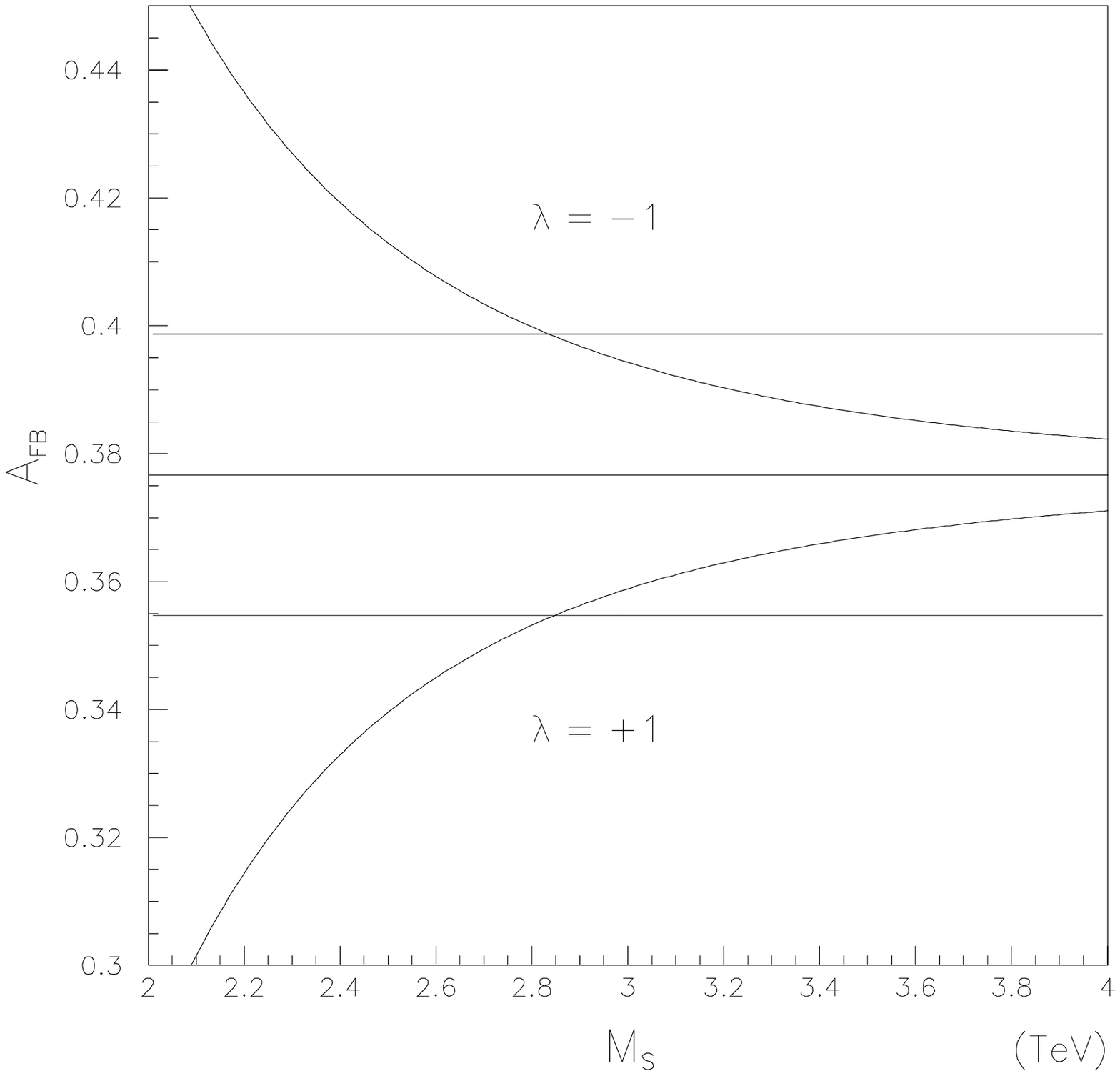,height=9cm,width=11cm}
\vspace{0.2cm}
\begin{center}
{\footnotesize
Fig. 2 : 
The forward-backward asymmetries with respect to $M_S$.
The straight lines denote the SM predictions at $2\,\sigma$ level.
}
\end{center}
\end{figure}

\section{Further Study : $\gamma \gamma$ collision}

Photon-photon collisions have been regarded
as one of the best alternatives of $e^+ e^-$ collisions at LC,
where high energy photon beams can be achieved through laser
back-scattering of the parent $e^+e^-$ beams.
The controllability of the laser and electron beam polarizations
provides good opportunities to probe new physics.
It is shown that
photon colliders are sensitive to the presence of large extra
dimensions, yielding higher low bound of the $M_S$
than any other collider \cite{gg}.
The role of polarizations, however,
has been studied only about the laser and parent electron beams.

In this section we study the large extra dimension effects on the
polarizations of top quark pair production at photon colliders,
including the top spin configuration
which shall be shown to provide a unique channel.
From the total cross sections in various top spin bases,
we will observe that there is no particular top spin basis
where the pattern of the spin configuration
is crucially modified
by the large extra dimension effects.
Instead a new observable,
the `top spin asymmetry', is to be introduced.
There exists a special spin basis
where this top spin asymmetry vanishes in the SM but has
substantial values with low scale quantum gravity effects.

For the process
\begin{equation}
\label{process}
\gamma(k_1,\lambda_1) + \gamma(k_2,\lambda_2)
\to t(p_1,\kappa_1) + \bar{t}(p_2, \kappa_2),
\end{equation}
there are two types of tree level diagrams in the SM,
$t$-channel and $u$-channel ones.
In Eq.~(\ref{process}), $\lambda_i$ and $\kappa_i$ $(i=1,2)$
denote the polarizations of photons and top quarks, respectively.
The effective theory of low scale quantum gravity allows an
$s$-channel diagram mediated by virtual KK graviton modes, 
of which the scattering amplitude is
\begin{eqnarray}
\label{amp}
{\cal M}_G &=& - \frac{2s\lambda}{M_S^4}\,\overline{u}(p_1)
   \left[ ( j\cdot \varepsilon_2) \rlap/{\varepsilon}_1
   +( j\cdot \varepsilon_1) \rlap/{\varepsilon}_2
\right.
\\ \nonumber 
&&~~~~~~~~ \left.  + 2 {\varepsilon}_1 \cdot {\varepsilon}_2
   \left(
        \frac{(j\cdot k_2) \rlap/k_1 + (j\cdot k_1) \rlap/k_2 }{s}-m_t
        \right)
\right] v(p_2)\,,
\end{eqnarray}
where $j=p_1-p_2$,
and $\varepsilon_i=\varepsilon(k_i)\,\,(i=1,2)$ are
the polarization vectors of initial photons.

When the top quark is emitted in the $z$-axis,
the polarization vectors of initial photons 
are given by, 
in the center of momentum (CM) frame with the Coulomb gauge,
\begin{equation}
\label{pol}
\varepsilon_1(\pm)=\varepsilon_2(\mp)=\frac{1}{\sqrt{2}}
\big( 0, \mp \cos\theta, -i, \mp\sin\theta\big),
\end{equation}
where $\theta$ is the scattering angle.
There is a relation\cite{CCIS}, usually being used in the squared amplitudes,
\bea
\label{ggpol}
\varepsilon^i(k,\lambda_1) \varepsilon^{*j}(k,\lambda_2)
&=&
\frac{1}{2}
\left[
(\delta^{i j} - \hat{k}^i  \hat{k}^j ) \delta_{\lambda_1\lambda_2}
-\frac{i}{2} (\lambda_1 +\lambda_2)\epsilon^{ijk} \hat{k}^k \right.
\\ \nonumber &&\quad
-\frac{1}{2}(\lambda_1 -\lambda_2)
\left\{
\hat{a}^i(\hat{\bf k} \times \hat{\bf a})^j 
+ \hat{a}^j(\hat{\bf k} \times \hat{\bf a})^i
\right\}
\\ \nonumber &&\quad
\left.
+\frac{1}{2}(\lambda_1\lambda_2-1)
\left\{
\hat{a}^i\hat{a}^j
-(\hat{\bf k} \times \hat{\bf a})^i (\hat{\bf k} \times \hat{\bf a})^j
\right\}
\right]
\,,
\eea
where $\hat{\bf a}$ is an arbitrary unit vector perpendicular to $\hat{\bf k}$,
and $i,j=1,2,3$.

Equation (\ref{ggpol}) can be applied even in the amplitude level.
In the process of Eq.~(\ref{process}),
two channels are possible according to the total angular momentum:
$J_z = 0 $ and $J_z = 2 $ states. 
In the $J_z =0$ states ($\gamma_R\gamma_R$ or $\gamma_L\gamma_L$),
two polarization vectors are related by
$\varepsilon_2 (+)= -\varepsilon_1 ^\ast (+)$,
as can be easily seen from Eq.~(\ref{pol}).
Then the $\varepsilon^\mu_1\varepsilon^\nu_2$
in Eq.~(\ref{amp}) can be replaced by
\be
\label{polpol}
\varepsilon^i_1(+1)
\varepsilon^j_2(+1)
=
-\varepsilon^i(k_1,+1) \varepsilon^{*j}(k_1,+1)
\,.
\ee
Note that only the first term in Eq.~(\ref{ggpol}) 
contributes to the scattering amplitude in Eq.~(\ref{amp})
since the last two terms vanish in the case of $\lambda_1=\lambda_2$,
and the anti-symmetric second term does not contribute to 
the amplitude symmetric under the exchange of two photons.
In the CM frame where $ j \cdot k_1=- j \cdot k_2 =-{\bf j}\cdot{\bf k_1}$
and $\bar{u}(p_1) \rlap/k_1 v(p_2) = -\bar{u}(p_1) \rlap/k_2 v(p_2) 
=\bar{u}(p_1) \gamma_i \, v(p_2) k_1^i$,
the scattering amplitude mediated by the KK modes
vanishes:
\be
\label{zero-amp}
{\mathcal M}^{RR}_G
=
\frac{2s\lambda}{M_S^4}
\left[
\bar{u}\gamma_i v j^j \left(
\delta^{ij}-\frac{k_1^i k_1^j}{k_1^0 k_1^0 } \right)
-\bar{u}\gamma_i v j^i 
+\frac{ \bar{u}\gamma_i v k_1^i ({\bf j}\cdot{\bf k}_1)}{k_1^0 k_1^0}
\right]
 =0
\,.
\ee
Thus the effects of low scale quantum gravity exist only in
the $J_z=2$ channels. 

The squared amplitudes for each top spin 
configuration with the large extra dimension effects are,
with $\beta=\sqrt{1-4m_t^2/s}$ \cite{leeggtop},
\begin{eqnarray}
\label{ampsq}
|{\cal M}|^2 (\gamma_R \gamma_L \rightarrow
t_\uparrow \overline{t}_\uparrow ~{\rm or}~ t_\downarrow
\overline{t}_\downarrow ) &=& \frac{1}{4}N_cs^2
\beta^2\sin^2\theta[D_t+D_u+D_s]^2F_1^2,
\nonumber \\
|{\cal M}|^2  (\gamma_R \gamma_L \rightarrow
t_\uparrow \overline{t}_\downarrow ~{\rm or}~ t_\downarrow
\overline{t}_\uparrow ) &=& \frac{1}{4}N_cs^2
\beta^2\sin^2\theta[D_t+D_u+D_s]^2(F_2\mp1)^2,
\nonumber \\ 
|{\cal M}|^2  (\gamma_R \gamma_R \rightarrow
t_\uparrow \overline{t}_\uparrow ~{\rm or}~ t_\downarrow
\overline{t}_\downarrow ) &=& \frac{1}{4}N_cs^2
(1-\beta^2)[D_t+D_u]^2(1\mp\beta\cos\xi)^2,
\nonumber \\
|{\cal M}|^2(\gamma_R \gamma_R \rightarrow
t_\uparrow \overline{t}_\downarrow ~{\rm or}~ t_\downarrow
\overline{t}_\uparrow ) &=& \frac{1}{4}N_cs^2
\beta^2(1-\beta^2)[D_t+D_u]^2\sin^2\xi,
\end{eqnarray}
where $N_c$ is the number of color, 
$D_{s,t,u}$ are the effective propagation factors defined by 
\begin{equation}
D_t=\frac{Q_t^2 e^2}{t-m_t^2},\quad
D_u=\frac{Q_t^2 e^2}{u-m_t^2},\quad
D_s=\frac{4s\lambda}{M_S^4},
\end{equation}
and $F_{1,2}$ are the spin configuration factors as
functions of the scattering angle, given by
\begin{eqnarray}
F_1&=&\sqrt{1-\beta^2}\sin\theta\cos\xi-\cos\theta\sin\xi, 
\nonumber \\
F_2&=&\sqrt{1-\beta^2}\sin\theta\sin\xi+\cos\theta\cos\xi,
\end{eqnarray}
with the angle $\xi$ introduced in the previous section,
which represents the top spin configuration.
The SM results are in agreement with those of Ref.~\cite{Hori}.
It has been noted that the processes
$\gamma_R\gamma_L \to (t_\uparrow \bar{t}
_\uparrow ~{\rm or}~ t_\downarrow\bar{t}_\downarrow)$
permit a special top spin
basis where the SM prediction vanishes.
The other amplitudes can be obtained by using the
$CP$ invariance such as 
\begin{equation}
\label{cp}
|{\cal M}_{RL\uparrow \uparrow}| = |{\cal M}_{LR\downarrow \downarrow}|,
\quad
|{\cal M}_{RR\uparrow \downarrow}| = |{\cal M}_{LL\downarrow \uparrow}|
,
\end{equation}
where e.g., the suffix $RL\uparrow\uparrow$ denotes the process
$\gamma_R\gamma_L\rightarrow t_\uparrow \bar{t}_\uparrow$. 

The squared amplitudes in Eq.~(\ref{ampsq})
imply important characteristic features of 
the $\gamma\gamma \to t \bar{t}$ process.
First the low scale quantum gravity effects contained in $D_s$
exist only in the $J_z=2$ cases,
as discussed before.
Second the effects in any top spin basis do not modify
the top spin configuration,
which is not the case at $e^+e^-$ colliders.

We briefly review the differential cross sections at practical
$\gamma\gamma$ colliders \cite{Ginzburg}.
From the head-on collisions between the laser
and energetic electron (or positron) beams,
high energy polarized photons are produced.
When $x$ denotes the fraction of the photon beam energy 
to initial electron beam energy, i.e., $x=E_\gamma/E$,
its maximum value is $x_{max}=z/(1+z)$
where $z=4E \omega_0/m_e^2$.
Here $E_\gamma$, $E$, $\omega_0$ are the photon, electron and laser beam 
energies, respectively. 
It is known that laser beam with too high energy would produce
$e^+e^-$ pair through collisions with the back-scattered photon beam,
reducing the $\gamma\gamma$ luminosity.
Thus the $z$ is optimized to be $2(1+\sqrt{2})$ which occurs
at the threshold for the electron pair production.
In the numerical analysis,
we consider the following cuts:
\begin{eqnarray}
&&-0.9 \le \cos \theta \le 0.9, \nonumber \\
&& \sqrt{0.4}\le x_{1(2)} \le x_{max}|_{z=2(1+\sqrt{2})}.
\end{eqnarray}

With given polarizations of the laser and parent electron beams, 
the differential cross section 
is,
taking into account of their Compton back-scattering,
\begin{eqnarray}
\label{conv}
\frac{d\sigma}{d\cos\theta}&=&\frac{1}{32\pi s_{ee}}
\int \int dx_1dx_2 \frac{f(x_1)f(x_2)}{x_1x_2} \\
&&\times \bigg[ \Big( \frac{1+\xi_2(x_1)\xi_2(x_2)}{2}\Big) 
\Big|{\cal M}_{J_z=0} \Big|^2 
+\Big( \frac{1-\xi_2(x_1)\xi_2(x_2)}{2}\Big) 
\Big|{\cal M}_{J_z=2} \Big|^2 \bigg], \nonumber
\end{eqnarray}
where
\begin{eqnarray}
\label{sum}
\Big|{\cal M}_{J_z=0} \Big|^2 &=& \frac{1}{2}\bigg[
\Big|{\cal M}_{RR}\Big|^2 + \Big|{\cal M}_{LL}\Big|^2 \bigg], 
\nonumber \\
\Big|{\cal M}_{J_z=2} \Big|^2 &=& \frac{1}{2}\bigg[
\Big|{\cal M}_{RL}\Big|^2 + \Big|{\cal M}_{LR}\Big|^2 \bigg]. 
\end{eqnarray}
The function $f(x)$ and $\xi_2(x)$ are the photon number density
and the averaged circular polarization of the
back-scattered photon beams, respectively\cite{davoudiasl,CCIS,choi}.
As $f(x)$ implies,
the back-scattered photons possess non-trivial energy spectrum,
which renders unavoidable finite mixing of the $J_z=2$ and $J_z=0$ states.
The $s_{ee}$ is the CM squared energy 
at $e^+ e^-$ collisions, related with that of
$\gamma\gamma$ collisions, $s$, by $s=x_1 x_2 s_{ee}$.
Since the $\gamma\gamma$ luminosity spectrum 
shows a narrow peak 
at around $x_{max}$
with an appropriate cut on the longitudinal momentum
of the backscattered photons \cite{Ginzburg},
the approximations such as $t=x_1 x_2 t_{ee}$ and $u=x_1 x_2 u_{ee}$
are to be employed \cite{davoudiasl,CCIS,choi}. 

\begin{figure}[t]
\epsfig{file=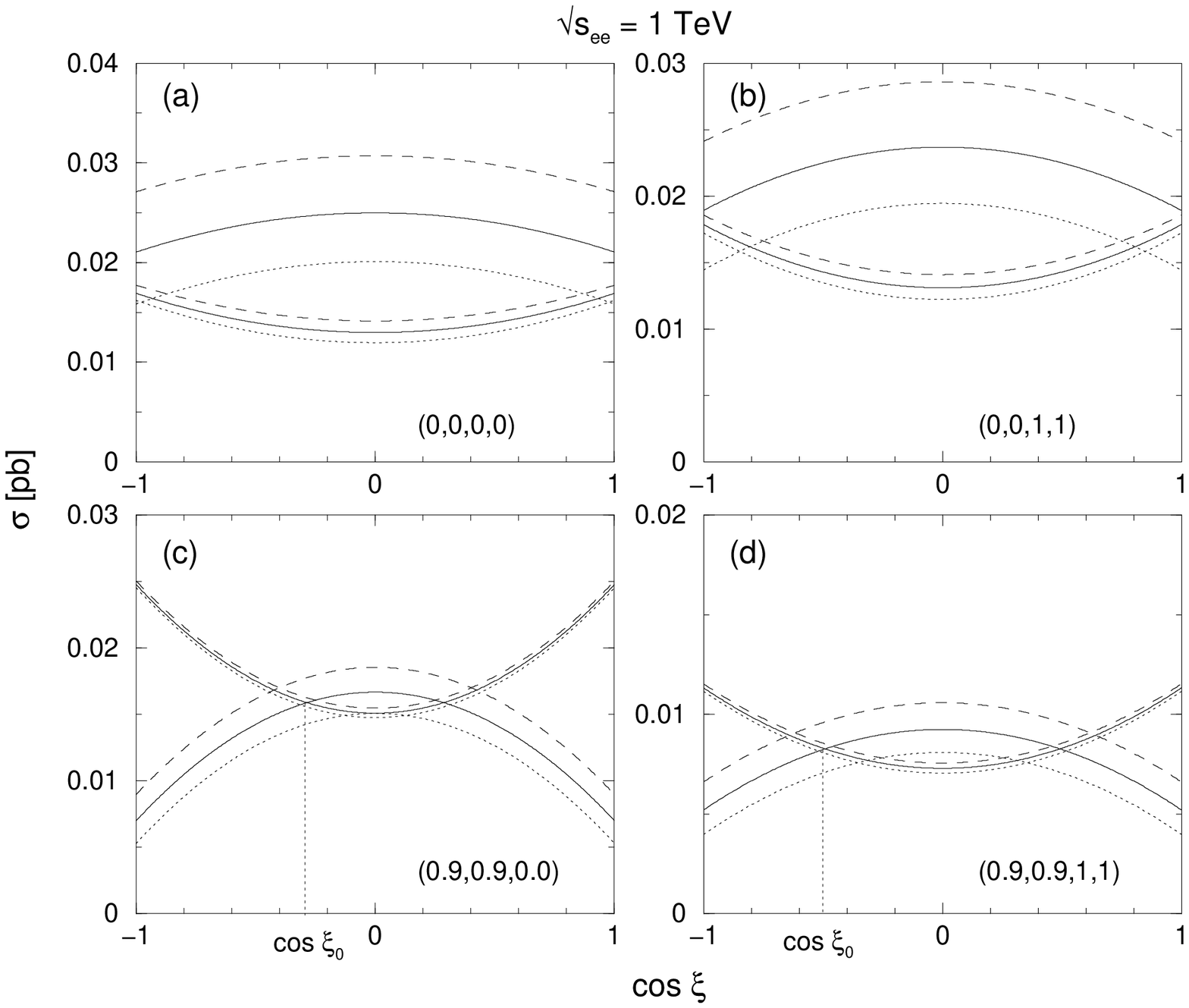,height=10cm,width=12cm}
\vspace{0.2cm}
\begin{center}
{\footnotesize 
Fig. 3 : Total cross sections as a function of $\cos \xi$: (a) with the
unpolarized beams, i.e., $(P_{e1}, P_{e2},P_{l1},P_{l2})=(0,0,0,0)$;
(b) $(0,0,1,1)$; (c) $(0.9,0.9,0,0)$;
(d) $(0.9,0.9,1,)$.
}
\end{center}
\end{figure}

\begin{figure}[t]
\epsfig{file=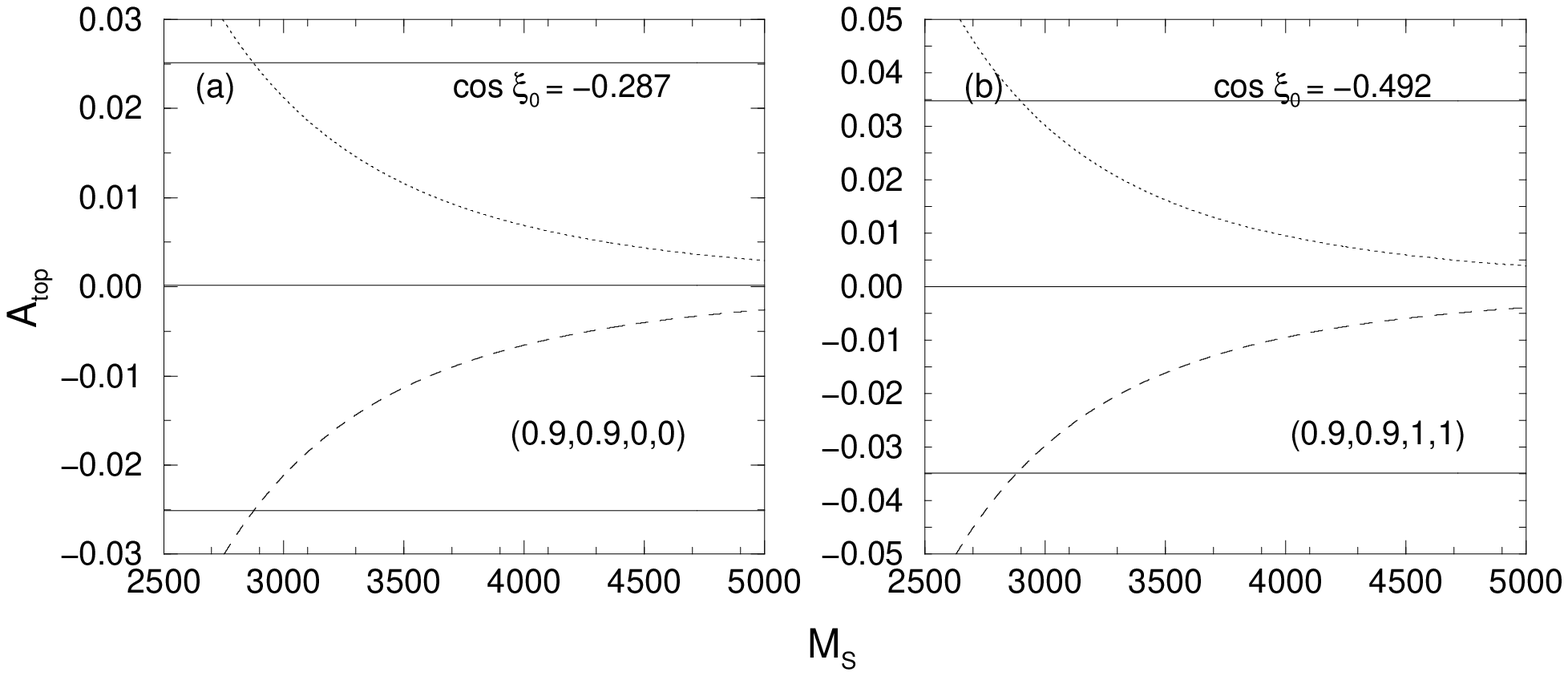,height=5cm,width=12cm}
\vspace{0.2cm}
\begin{center}
{\footnotesize 
Fig. 4 : 
The optimal top spin asymmetries with respect to $M_S$
with unpolarized and/or polarized initial beams.
The solid lines denote $2\,\sigma$ level predictions of the SM.
The dotted and dashed lines incorporate low scale quantum gravity effects
in the cases of $\lambda = 1$ and $\lambda = -1$, respectively.
}
\end{center}
\end{figure}

Top spin information with $CP$ invariance 
permits only two independent cross sections,
\begin{equation}
\sigma_{\uparrow\uparrow} (= \sigma_{\downarrow\downarrow}) ,~~  
\sigma_{\uparrow\downarrow} (= \sigma_{\downarrow\uparrow}), 
\end{equation}
which can be easily seen from Eqs.~(\ref{cp}) and (\ref{sum}).
In Fig.~3, we present $\sigma_{\uparrow\uparrow}$ 
and $\sigma_{\uparrow\downarrow}$ with respect to
$\cos\xi$ at $\sqrt{s_{ee}}=1$ TeV
with unpolarized and/or polarized initial beams. 
The polarizations of the laser and electron beams
are allowed as 100\% and 90\%, respectively.
The $M_S$ is set to be 2.5 TeV. 
The convex curves denote the 
$(t_\uparrow \bar{t}_\downarrow)$ case, while
the concave ones do the $(t_\uparrow \bar{t}_\uparrow)$ case.
The solid lines describe the SM
results, and the dotted and dashed lines describe the results 
including low scale quantum gravity effects
with $\lambda=1$ and $\lambda=-1$, respectively.
Each value of $\cos\xi$ indicates different spin basis
of the top quark and anti-quark.

There are several interesting characteristics of 
photon colliders in producing a top quark pair.
As can be first seen in Fig.~3, 
there is no special $\cos\xi$ at which
one top spin configuration dominates over the other, as discussed before.
The quantum gravity effects manifest themselves
as being factored out from the top spin configuration, 
i.e., independent of $\cos\xi$.
And irrespective of the polarizations of the laser and electron beams,
the effects are larger in the cases of $t_\uparrow \bar{t}_\downarrow$
than in those of $t_\uparrow \bar{t}_\uparrow$.
Equation (\ref{ampsq}) suggests its reason.
For instance, with the unpolarized beams,
the ratio $|{\mathcal M}_{RL}/{\mathcal M}_{RR}|^2$ hints 
the magnitude of the quantum gravity effects,
since the $J_z=2$ states possess the effects but the $J_z=0$ states do not.
Hence the top quark pair production channel with the 
$(t_\uparrow \bar{t}_\downarrow)$ is effective to probe
the quantum gravity effects.
Finally we observe that the results with the polarized
electron beams are of great interest and importance:
There exist two values of $\cos\xi$, i.e., 
two special spin bases where the equalities of
$\sigma_{\uparrow \uparrow } = \sigma_{\uparrow \downarrow}$ hold true.
Considering a new observable, top spin asymmetry $A^{top}$ \cite{leeggtop},
\begin{equation}
A^{top} \equiv {{ \sigma_{\uparrow\downarrow} -\sigma_{\uparrow\uparrow}}
\over{ \sigma_{\uparrow\downarrow} +\sigma_{\uparrow\uparrow}}},
\end{equation}
we suggest an optimal top spin 
basis defined by $\xi_0$ which satisfies
\begin{equation}
\sigma_{\uparrow\downarrow} ^{SM} ( \xi_0) =
\sigma_{\uparrow\uparrow} ^{SM} ( \xi_0)
\,.
\end{equation}
The definition guarantees that the observable
$A^{top}(\xi_0)$ vanishes in the SM
while retains non-vanishing values 
with the low scale quantum gravity effects.
The $A^{top}(\xi_0)$ as a function of $M_S$ is plotted
in Fig.~4.  
The solid lines indicate the SM predictions
at $2\sigma$ level,
while the dotted and dashed lines include
the quantum gravity effects of $\lambda=1$ and
$\lambda=-1$, respectively.
The case of $M_S=2.5$ TeV at $\sqrt{s_{ee}}=1$ TeV
causes about $A^{top}(\xi_0)$ of 6\%.

The $A^{top}(\xi_0)$ observables yield lower bounds
on the string scale $M_S$, which can be read off from Fig. 4.
In Table I, we summarize other lower bounds of the $M_S$,
which can be obtained by using various combination
of the polarizations of the laser and electron beams,
and by measuring the top spin configuration.
We have employed the luminosity ${\mathcal L} = 200$ fb$^{-1}/$yr.
As discussed before, the top spin state 
$(t_\uparrow \bar{t}_\downarrow)$  can constrain most strictly
the $M_S$, 
though the enhancement of the $M_S$ bound 
is not very significant.
This is mainly due to the statistical disadvantage of 
the top spin measurements.

\vskip 0.5cm
\noindent
\begin{center}
\begin{tabular}{|c|cc|cc|cc|}
\hline
&
\multicolumn{2}{|c|}{$\sigma_{tot}$}
&
\multicolumn{2}{|c|}{$\sigma_{\uparrow\uparrow}$}
&
\multicolumn{2}{|c|}{$\sigma_{\uparrow\downarrow}$}
\\ 
$(P_{e1,e2,l1,l2})$ & ~$\lambda=1 $~ & ~$\lambda=-1 $~ &
~  $\lambda=1 $ & ~$\lambda=-1 $~ &
~  $\lambda=1 $~ &~ $\lambda=-1 $ ~\\
\hline
(0,0,0,0) & 4.1 & 4.1 & 2.6 & 2.8 & 4.3 & 4.3 \\
(0,0,1,1) & 4.0 & 4.0 & 2.5 & 2.7 & 4.2 & 4.2 \\
(0.9,0.9,0,0) & 3.2 & 3.2 & 1.7 & 2.0 & 3.7 & 3.7 \\ 
(0.9,0.9,1,1) & 3.2 & 3.2 & 1.8 & 2.0 & 3.5 & 3.6  \\
\hline
\end{tabular}
\end{center}
\vskip 0.3cm
{Table~I}. The photon collider bounds of $M_S$ in TeV at $2\sigma$
level according to the polarizations of the laser, the parent 
electron beam, and top spin configurations.
\vskip 0.5cm

\section{Concluding Remarks}

We have studied the spin configuration
of the top quark pair at the $e^+ e^- \to t \bar{t}$
and $\gamma \gamma \to t \bar{t}$ processes
with the large extra dimension.
In the framework of the off-diagonal basis where
the cross sections of the like-spin states of the top quark pair
vanish in the SM, the left-handed (right-handed)
electron beam almost completely prefer 
the spin configuration of the top and anti-top spins as Up-Down (Down-Up)
at the $e^+ e^-$ collision.
The presence of large extra dimensions modifies these features
significantly at high energies
by yielding non-zero cross sections for the like-spin states. 
In addition, it is shown that the forward-backward asymmetry
is very effective to probe the new physics scale $M_S$
and to determine the sign of the quantum gravity corrections.

From the polarized scattering amplitudes of 
$\gamma \gamma \to t \bar{t}$ process in a general top spin basis,
it has been shown that only the $J_z=2$ states receive the effects
which do not affect the top spin configuration.
The non-trivial energy spectrum of the Compton back-scattered photons
leads to finite mixing of the $J_z=0$ and $J_z=2$ states,
as well as permitting low scale quantum gravity to
play a role in modifying the top spin configuration.
This practical mixing of the $J_z=0$ and $J_z=2$ states
prevents us to define a top spin basis
in which one of the final spin configurations is dominant 
and the other is suppressed at the $\gamma \gamma$ collision.
Instead we have noticed that a special spin basis,
especially with the polarized parent electron beams,
exist such that $\sigma^{SM}(t_\uparrow\bar{t}_\downarrow)
=\sigma^{SM}(t_\uparrow\bar{t}_\uparrow)$ in the SM.
Accordingly introduced is the top spin asymmetry $A^{top}$
of which the SM predictions vanish in this optimal top spin basis,
which can be one of the most effective tools in this case.

We conclude that the spin configuration of the top pair 
is effective to explore the large extra dimension effects
at the future $e^+ e^-$ Linear Colliders.

\end{document}